\documentclass[aps,prd,twocolumn,showpacs,amsmath,amssymb,nofootinbib,eqsecnum, preprintnumbers]{revtex4-1}

\usepackage{color}

\newcommand{\ph}{\varphi}
\newcommand{\tht}{\vartheta}
\newcommand{\eps}{\varepsilon}
\newcommand{\one}{\textsc{i}}

\newcommand{\dm}{n}

\newcommand{\be}{\begin{equation}}
\newcommand{\ee}{\end{equation}}
\newcommand{\ba}{\begin{eqnarray}}
\newcommand{\ea}{\end{eqnarray}}

\newcommand{\beq}{\begin{equation}}
\newcommand{\eeq}{\end{equation}}
\newcommand{\beqa}{\begin{eqnarray}}
\newcommand{\eeqa}{\end{eqnarray}}
\newcommand{\nn}{\nonumber}

\newcommand{\hook}{\raisebox{-0.35ex}{\makebox[0.6em][r]
{\scriptsize $-$}}\hspace{-0.15em}\raisebox{0.25ex}{\makebox[0.4em][l]{\tiny
 $|$}}}
\newcommand{\cwedge}[1]{\mathop{\wedge}_{{}^{#1}} }

\newcommand{\even}{{\mathrm{e}}}
\newcommand{\odd}{{\mathrm{o}}}

\newcommand{\fiber}{\boldsymbol}

\newcommand{\cd}[1]{\frac{\eth}{\partial{#1}}}
\newcommand{\cv}[1]{{\partial}_{#1}}

\newcommand{\Ric}{{\mathrm{Ric}}}
\newcommand{\LCT}{z}

\newcommand{\lst}[1]{\langle{#1}\rangle}

\newcommand{\ef}{E}

\newcommand{\PKV}{\xi}
\newcommand{\KV}[1]{\xi_{(#1)}}
\newcommand{\KVf}[1]{\xi^{(#1)}}
\newcommand{\KT}[1]{k^{(#1)}}
\newcommand{\CCKY}[1]{h^{(#1)}}
\newcommand{\KY}[1]{f^{(#1)}}

\newcommand{\A}[1]{A^{(#1)}}

\newcommand{\Bo}[1]{\mathcal{B}^{(#1)}}
\newcommand{\Vo}{\mathcal{V}}

\newcommand{\psc}[1]{\Psi_{#1}}
\newcommand{\chc}[1]{\mathcal{X}_{#1}}
\newcommand{\psf}[1]{\tilde\Psi_{#1}}
\newcommand{\chf}[1]{\tilde{\mathcal{X}}_{#1}}

\newcommand{\clfm}[1]{{\slash\mspace{-9.5mu}{#1}}}

\newcommand{\hdotp}[1]{h_{(#1)}}
\newcommand{\omh}[1]{\omega_{(#1)}}

\begin{document}
\title{Electron in higher-dimensional weakly charged rotating black hole spacetimes}

\author{Marco Cariglia}
\email{marco@iceb.ufop.br}
\affiliation{Universidade Federal de Ouro Preto, ICEB, Departamento de F\'isica.\\  Campus Morro do Cruzeiro, Morro do Cruzeiro,
35400-000 - Ouro Preto, MG - Brasil}

\author{Valeri P. Frolov}
\email{vfrolov@ualberta.ca}
\affiliation{Theoretical Physics Institute,
University of Alberta,\\
Edmonton, Alberta, Canada T6G 2G7}

\author{Pavel Krtou\v{s}}
\email{Pavel.Krtous@utf.mff.cuni.cz}
\affiliation{Institute of Theoretical Physics,
Faculty of Mathematics and Physics, Charles University in Prague,
V~Hole\v{s}ovi\v{c}k\'ach 2, Prague, Czech Republic}

\author{David Kubiz\v n\'ak}
\email{dkubiznak@perimeterinstitute.ca}
\affiliation{Perimeter Institute, 31 Caroline St. N. Waterloo Ontario, N2L 2Y5, Canada}

\date{November 17, 2012}  

\begin{abstract}
We demonstrate separability of the Dirac equation in weakly charged rotating black hole spacetimes in all dimensions. The electromagnetic field of the black hole is described by a test field approximation, with vector potential proportional to the primary Killing vector field. It is shown that the demonstrated separability can be intrinsically characterized by the existence of a complete set of mutually commuting first order symmetry operators generated from the principal Killing--Yano tensor. The presented results generalize the results on integrability of charged particle motion and separability of charged scalar field studied in \cite{FrolovKrtous:2011}.
\end{abstract}

\pacs{04.50.-h, 04.50.Gh, 04.70.Bw, 04.20.Jb}
\preprint{pi-stronggrv-303}

\maketitle
\section{Introduction}\label{sec1}
In our  four-dimensional universe `real' astrophysical black holes cannot carry large electric charge. In the  presence of  surrounding plasma, flux of the plasma particles with the opposite charge rapidly reduces the original charge of the black hole and makes it small. In the absence of monopoles a black hole cannot also carry a magnetic charge. However, in the presence of accreting disk, magnetic field can exist in the black hole vicinity. Moreover it is plausible that it plays an important role in a black hole activity. The electric and magnetic fields $E$ and $B$ carry energy density $\varepsilon \sim |E|^2+|B|^2$, but, in a realistic situation its contribution to the curvature $G\varepsilon/c^4$ is much smaller than the characteristic curvature of the spacetime near the horizon ${\sim}  (\text{grav.radius})^{-2}$. This means that the distortion of the background gravitational field is small and can be neglected. We call such electric and magnetic fields {\em weak}. This does not mean at all that their role in the physical processes in the black hole vicinity is also negligibly small. The reason is that for charged particles the relative weakness of the electromagnetic field  can be compensated by a large value of the charge-to-mass ratio, which, for example, for electrons is $e/m\approx 5.2728 \times 10^{17} g^{-1/2}cm^{3/2}s^{-1}$. In other words, the motion of charged particles in weakly charged and/or weakly magnetized black holes might be quite different from the motion of neutral ones.

For example, the innermost stable orbit of a charged particle moving in the equatorial plane of a weakly magnetized black hole can be very close to the horizon, see, e.g., \cite{FrolovShoom:2010}. In consequence, as it was shown recently,  weakly magnetized black holes may play a role of particle accelerators \cite{Frolov:2012,IgataEtal:2012}.

The study of weakly charged and/or magnetized black hole is greatly simplified, since the electromagnetic field can be considered as a test one. The electromagnetic field $A$ in the Lorentz gauge $\nabla_{\!a}A^a=0$ obeys the equation
\be \label{eq:Maxwell_curved}
\Box  A_{a} {\color{red}}- \Ric^{b}{}_{a} A_b=0\, .
\ee
If the spacetime possesses a Killing vector $\xi$, it is easy to show that it satisfies the equation
\be \label{eq:Killing_curved}
\Box  \xi_{a} {\color{red}}+ \Ric^{b}{}_{a} \xi_{b}=0\, .
\ee
Hence in a Ricci-flat spacetime $\xi$ is a potential for a special test electromagnetic field. This observation by \cite{Papapetrou:1966, Wald:1974} is well known and often used. Let us emphasize that this property  is valid in a spacetime with arbitrary number of dimensions. A well known example is the Kerr spacetime with two Killing vectors, $\xi_{(t)}$ and $\xi_{(\ph)}$, generating time translation and axial rotation symmetries.
The corresponding test Maxwell fields describe respectively a weak electric field of a {\em test charge} inside
the black hole, and the axisymmetric stationary {\em magnetic field} near the black hole
which is homogeneous at the infinity and directed along the axis of rotation.
Evidently, this construction can be easily generalized to the Myers--Perry spacetimes
in an arbitrary number of dimensions, e.g.,  \cite{AlievFrolov:2004, Aliev:2006b, Aliev:2007, Krtous:2007}.

In spite of the fact that the relation between special test fields and the Killing vectors generating them is of a general nature,   the character of the motion of a charged particle in the presence of such a field is quite different for different fields. To illustrate this, let us consider again the case of 4D Kerr spacetime. The geodesic equations for particles in the Kerr metric are {\em completely integrable} \cite{Carter:1968pr}.  This property is preserved for the motion of electrically charged particles in a weakly charged black hole. (In fact, it is valid for the Kerr--Newman spacetime, where the charge of a black hole is not weak.) The case of a weakly magnetized Kerr black hole is quite different. Particles moving in the equatorial plane are still completely integrable, while the `fourth' (Carter's) integral of motion is absent for the generic motion out of the symmetry plane. As a result, in a general case the motion of a charged particle in weakly magnetized black holes is chaotic.

It is interesting, that this is a generic feature of higher dimensional rotating black holes as well. Namely, it was shown in \cite{FrolovKrtous:2011} that the charged particle equations in higher dimensional rotating black holes with a test electric field proportional to the {\em primary} (timelike) {\em Killing vector field} are completely integrable---the properties of such spacetimes are similar to the properties of a weakly charged Kerr spacetime. However, complete integrability is destroyed in the presence of a test electromagnetic field generated by {\em secondary} (rotational) {\em Killing vector fields}. (This is an analogue of weakly magnetized black holes.) It was shown that the Klein--Gordon and Hamilton--Jacobi equations for such a system allow complete separation of variables only for the primary Killing test field \cite{FrolovKrtous:2011}. The latter stems from the existence of the Lax tensor for a charged particle---discussed recently in \cite{CarigliaEtal:2012}.

In this paper we study the properties of a charged Dirac equation in higher dimensional rotating black hole spacetimes in the presence of a test electromagnetic field derived from various Killing vector fields, and demonstrate, that the structure outlined above for particles and scalars remains similar. Namely, for the weakly charged black holes the Dirac equation preserves its remarkable property and allows a complete separation of variables. Whereas we show that this is not possible for a magnetized black hole.

The paper is organized as follows. In the next section we review the charged Dirac equation in curved space and discuss its first order symmetry operators.
In Sec.~\ref{sec3} we gather the information about weakly charged and magnetized rotating black holes in any dimension. In Sec.~\ref{sec4} we demonstrate the existence of a complete set of first order mutually commuting operators for the charged Dirac equation in a weakly charged black hole spacetime. The explicit representation of these operators is found in Sec.~\ref{sec5}. Separability of the charged Dirac equation in weakly charged black hole spacetimes is demonstrated in Sec.~\ref{sec6}. We conclude in Sec.~VII. Appendix~\ref{appA} displays a derivation of the important identity which is crucial for showing the mutual commutation of the symmetry operators in the set. Appendix~\ref{appB} offers an alternative proof of this property employing the explicit representation of these operators.
In many places the paper closely follows \cite{CarigliaEtal:2011b} where similar results
were proved for the standard (uncharged) Dirac equation in higher-dimensional Kerr-NUT-AdS spacetimes.

\section{Charged Dirac equation in curved space}\label{sec2}

\subsection{Dirac bundle}
In what follows we parametrize a  dimension of the spacetime as
\be
\dm = 2N + \eps\,,
\ee
with $\eps =0, 1$ corresponding to the even, odd dimension, respectively. The Dirac bundle ${\fiber{D}M}$ has fiber dimension ${2^N}$.
It is connected with the tangent bundle ${\fiber{T}M}$ of the spacetime manifold ${M}$ through the abstract gamma matrices ${\gamma^a\in \fiber{T}M\otimes\fiber{D}^1_1 M}$,
which satisfy
\begin{equation}\label{gamgamg}
    \gamma^a\gamma^b+\gamma^b\gamma^a=2\,g^{ab}\;.
\end{equation}
They generate an irreducible representation of the abstract Clifford algebra on the Dirac bundle. All linear combinations of products of the abstract gamma matrices (with spacetime indices contracted) form the Clifford bundle ${\fiber{Cl}M}$, which is thus identified with the space ${\fiber{D}^1_1M}$ of all linear operators on the Dirac bundle. The Clifford multiplication (`matrix multiplication') is denoted by juxtaposition of the Clifford objects. The gamma matrices also provide the Clifford map ${\gamma_*}$, which is an isomorphism between the exterior algebra ${\fiber{\Lambda}M}$ and the Clifford bundle ${\fiber{Cl}M}$,
\begin{equation}\label{cleaiso}
    \clfm\omega \equiv \gamma_* \omega
      \equiv \sum_p \frac1{p!}\, (\omega_p)_{a_1\dots a_p}\gamma^{a_1\dots a_p}\;.
\end{equation}
Here, ${\omega = \sum_p  \omega_p\in\fiber{\Lambda}M}$ is an inhomogeneous form, $\omega_p$ its $p$-form part, $\omega_p\in \fiber{\Lambda}^p M$, and ${\gamma^{a_1\dots a_p} = \gamma^{[a_1}\cdots\gamma^{a_p]}}$.
For future use we also define an operator $\pi$ as $\pi\omega=\sum_p p\,\omega_p$.

We denote the standard (uncharged) Dirac operator both in the exterior bundle and the Dirac bundle as ${D}$
\be\label{DO}
D = e^a\nabla_a\;,\qquad D=\gamma_* e^a\nabla_a=\gamma^a\nabla_a\,.
\ee
Here ${e^a\in\fiber{T}M\otimes\fiber{\Lambda} M}$ is a counterpart of ${\gamma_a}$ in the exterior algebra and $\nabla$ denotes the spinorial covariant derivative.
We shall also denote by ${X_a}$ the object dual to ${e^a}$ and by
${\displaystyle\cwedge{k}}$
a contracted wedge product, defined inductively as \cite{HouriEtal:2010a, KubiznakEtal:2010}
\be
{\alpha}
\cwedge{0} {\beta} ={\alpha}
\wedge {\beta}\,, \quad {\alpha}
\cwedge{k} {\beta} = ({X}^a \hook {\alpha}) \cwedge{k-1} ({X}_a
\hook {\beta})\,,
\ee
where the `hook' operator $\hook$ corresponds to the inner derivative. The reader is also referred to the Appendix of \cite{CarigliaEtal:2011a} for more details on the notation.

\subsection{Killing--Yano tensors}
A {\em Killing--Yano} (KY) tensor ${f}$  \cite{Yano:1952} is a $p$-form on the spacetime, the covariant derivative of which is completely determined by its antisymmetric part, i.e., by its exterior derivative ${df}$:
\begin{equation}\label{KYdef}
  \nabla_{\!a} f_{a_1\dots a_r} = \nabla_{\![a} f_{a_1\dots a_r]}
    = {\textstyle\frac1{p{+}1}} (df)_{aa_1\dots a_p}\;.
\end{equation}
A {\em closed conformal Killing--Yano} (CCKY) tensor $h$ is a \mbox{$p$-form} on the spacetime, the covariant derivative of which is entirely determined by its divergence ${\xi}$:
\begin{equation}\label{CCKYdef}
\begin{gathered}
  \nabla_{\!a} h_{a_1a_2\dots a_r} = r\, g_{a[a_1} \xi_{a_2\dots a_r]}\;,\\
  \xi_{a_2\dots a_r} = {\textstyle\frac{1}{\dm-r+1}} \nabla_{\!a} h^a{}_{a_2\dots a_r}\;.
\end{gathered}
\end{equation}
KY and CCKY tensors are related to each other through the Hodge duality: the Hodge dual of a KY form is a CCKY form and vice versa. As we shall see, both these tensors play a crucial role in the study of symmetry operators of the Dirac operator.

It is also useful to introduce the notion of a {\em principal Killing--Yano} (PKY) tensor. We define this to be a non-degenerate (with maximal matrix rank and functionally independent eigenvalues) rank-2 CCKY tensor \cite{KrtousEtal:2007jhep}. We shall see in Sec.~\ref{sec3} that this tensor exists in rotating black hole spacetimes in all dimensions. It generates a Killing vector $\PKV$,
\be\label{primary}
\PKV_a=\frac{1}{\dm-1}\nabla_{\!b} h^b{}_a\,,
\ee
which we call the {\em primary Killing vector}, as well as,  a whole set of
{\em secondary Killing vectors}, see Sec.~\ref{sec3}.

\subsection{Symmetry operators}
We shall now write down the most general first-order operators commuting with the $U(1)$ charged Dirac operator. The results follow from the general discussion in \cite{KubiznakEtal:2010} where the symmetry operators of a Dirac operator coupled to an arbitrary flux were studied (for special cases see also \cite{BennCharlton:1997} and  \cite{AcikEtal:2008c}).

Let ${\cal D}$ be the Dirac operator for a charged particle in the $U(1)$ potential $A$,
\be\label{chargedD}
{\cal D}=D-iA=e^a{\cal D}_a\,,\qquad {\cal D}=\gamma^a {\cal D}_a \,,
\ee
where we have introduced the $U(1)$ covariant spinorial derivative ${\cal D}_a = \nabla_a - i A_a$.

We first consider the case when the background metric admits a Killing vector equivalent to a Killing--Yano \mbox{1-form}~ $\kappa$. Then, the following operator:
\be\label{k}
K_\kappa= X^a\hook \kappa \nabla_{\!a}
       + \frac{1}{4}d\kappa - i A\cwedge{1}\kappa+i\ph\,,
\ee
where $\ph$ is an arbitrary function satisfying the `{\em anomaly condition}'
\be\label{anomalyk}
d\ph+(dA)\cwedge{1}\kappa=0\,,
\ee
commutes with the Dirac operator ${\cal D}$, \eqref{chargedD}. Given the potential $A$, one may try to find such $\ph$ so that the anomaly vanishes. Two possible choices are obvious (though not necessary): i) One chooses $\ph=0$ in which case the operator and the anomaly condition become
\begin{equation}
\begin{gathered}\label{k2}
K_\kappa= X^a\hook \kappa \, {\cal D}_{\!a}
       + \frac{1}{4}d\kappa \,,\\
0=\kappa\cwedge{1}dA\,.
\end{gathered}
\end{equation}
We shall see that a similar condition is required for higher-rank Killing--Yano tensors as well. ii) One chooses $\ph={\displaystyle A \cwedge{1}\kappa}$. In this case the operator \eqref{k} and the anomaly \eqref{anomalyk} simplify to
\begin{equation}
\begin{gathered}\label{Kk}
K_\kappa= X^a\hook \kappa \nabla_{\!a}
       + \frac{1}{4}d\kappa\,,\\
0={\cal L}_{\kappa^\sharp} A\,,
\end{gathered}
\end{equation}
where ${\cal L}_{\kappa^\sharp}$ is the Lie derivative along the Killing vector $\kappa^\sharp$. Hence, if $A$ respects the symmetry of $\kappa$, the operator \eqref{Kk} defines a symmetry operator (i.e., an operator commuting with ${\cal D}$) for the charged Dirac equation. In particular, as we shall see in Sec.~\ref{sec3}, this property is satisfied in weakly charged and magnetized rotating black hole spacetimes for $A$ associated with any Killing field, 
as all of them do mutually commute.

Since any Killing vector is Hodge dual to an \mbox{$(\dm{-}1)$-rank} CCKY tensor we do not have to  discuss the case of CCKY tensors of this rank anymore.

Having dealt with these special cases separately, we may now formulate the following result:
The most general (different from that discussed above) first-order operator ${S}$ which commutes with the Dirac operator ${\cal D}$, ${[{\cal D},S]=0}$, splits into the following Clifford even and odd parts:
\be\label{Scom}
  S=S_\even+S_\odd\,,
\ee
where
\ba
S_\even&=& L_{f_\odd} \equiv X^a\hook f_\odd \, {\cal D}_{\!a}
       + \frac{\pi-1}{2\pi}\,d f_\odd\,,\label{KdefA}   \\
S_\odd&=& M_{h_\even} \equiv e^a\wedge h_\even \, {\cal D}_{\!a}
       - \frac{\dm-\pi-1}{2(\dm-\pi)}\,\delta h_\even \;, \quad\label{MdefA}
\ea
with ${f_\odd}$ being an inhomogeneous odd (at least rank 3) KY form, and
${h_\even}$ being an inhomogeneous even (at most rank-($\dm{-}2$)) CCKY form, subject to the condition
\be\label{cond}
(dA)\cwedge{1} f_\odd=0\,,\quad
(dA)\cwedge{1} h_\even=0\,.
\ee
Note that in odd dimensions the operators $L$ and $M$ are related by the Hodge duality. However, this is not true in even dimensions, where these operators are necessarily independent.

Indeed, in an odd number of spacetime dimensions the Hodge duality of Killing--Yano tensors translates into the corresponding relation of symmetry operators $L$ and $M$.
Namely, let $\LCT$ be the Levi-Civita $\dm$-form satisfying
$\LCT_{a_1\dots a_n} \LCT^{a_1\dots a_n}= \dm !$.\footnote{%
Note that $\gamma_* (\LCT)$ is the ordered product of all $\dm$ gamma matrices and in odd dimensions it is proportional to the unit matrix. See also formula \eqref{gammaz}.
}
Then, in odd dimensions, the Hodge dual of a $p$-form $\omega$ can be written as
\be   \label{eq:HodgeDual}
*\omega = (-1)^{[\frac{p}{2}]} \,\LCT\, \omega \, ,
\ee
and for the operators of type $L$ and $M$ it holds that
\be  \label{eq:M_to_K}
L_{\LCT h}=\LCT\, M_h\,,\quad
M_{\LCT f} = \LCT\, L_{\!f} \, ,
\ee
where $f$ is an odd KY form and $h$ an even CCKY form. In particular, this is also valid for Killing vectors and dual CCKY $(\dm{-}1)$-forms.

%
The anomaly conditions \eqref{cond} are also related by the Hodge duality, thanks to the identity
\begin{equation}\label{hodgeFalph}
    F\cwedge{1}(*\alpha) = *(F\cwedge{1}\alpha)\,,
\end{equation}
which holds for any 2-form ${F}$ and an arbitrary form ${\alpha}$.

\section{Weakly charged rotating black hole spacetimes}\label{sec3}
In what follows we shall study the Dirac equation in the vicinity of the weakly charged and weakly magnetized rotating black hole spacetimes in all dimensions. Namely, as a background metric we shall use a `{\em canonical spacetime'}, which is the most general spacetime admitting the PKY tensor \cite{HouriEtal:2007, KrtousEtal:2008}. When vacuum Einstein equations are imposed such a spacetime describes the most general known rotating Kerr-NUT black hole discovered in \cite{ChenEtal:2006cqg}. The metric can be equipped with a weak electromagnetic field associated with $(N+\eps)$ isometries present in the spacetime. Among them the one associated with the primary Killing vector plays a special role. The background metric with vector potential proportional to this Killing field describes a weakly charged rotating black hole. We shall see that in this case the charged Dirac equation separates and the separability can be characterized by the existence of a complete set of mutually commuting first order symmetry operators. On the other hand, for weakly magnetized black holes, which are described by a vector potential associated with the secondary Killing vectors, the first order operators do not mutually commmute and the Dirac equation cannot be (at least in the canonical frame) separated. It is very likely that in this case
the integrability is lost and the Dirac equation does not separate in any frame.

\subsection{Canonical spacetimes}
The most general metric admitting the PKY tensor, also denoted as a canonical metric, writes in canonical coordinates as
\cite{HouriEtal:2007, KrtousEtal:2008}
\be  \label{metric}
{g}= \sum_{\mu=1}^N\biggl[ \frac{d x_{\mu}^{\;\,2}}{Q_\mu}
  +Q_\mu\Bigl(\,\sum_{j=0}^{N-1} \A{j}_{\mu}d\psi_j \Bigr)^{\!2}  \biggr]  + \eps S \Bigl(\,\sum_{j=0}^N \A{j}d\psi_j \Bigr)^{\!2},
\ee
where
\ba
Q_\mu&=&\frac{X_\mu(x_\mu)}{U_\mu}\,,\quad U_{\mu}=\prod\limits_{\nu\ne\mu} (x_{\nu}^2-x_{\mu}^2)  \;,\quad S = \frac{-c}{\A{N}} \, ,\quad\label{eq:UandS_def}\nonumber\\
\A{k}_{\mu}&=&\hspace{-5mm}\!\!
    \sum\limits_{\substack{\nu_1,\dots,\nu_k\\\nu_1<\dots<\nu_k,\;\nu_i\ne\mu}}\!\!\!\!\!\!\!\!\!\!
    x^2_{\nu_1}\cdots\, x^2_{\nu_k}\;,\ \
\A{j} = \hspace{-5mm} \sum\limits_{\substack{\nu_1,\dots,\nu_k\\\nu_1<\dots<\nu_k}}\!\!\!\!\!\!
    x^2_{\nu_1}\cdots\, x^2_{\nu_k}\; .\label{eq:A_def}\quad
\ea
Coordinates $x_\mu\, (\mu=1,\dots,N)$ stand for the (Wick rotated) radial coordinate and longitudinal angles. Killing
coordinates $\psi_k\; (k=0,\dots,N-1 +\eps)$ denote time and azimuthal angles associated with Killing vectors
${\KV{k}}$
\begin{equation}\label{KV}
\KV{k}=\cv{\psi_k}\;,\quad
\KVf{k}\equiv (\cv{\psi_k})^\flat\,.
\end{equation}
At points with ${x_\mu=x_\nu}$ with ${\mu\neq\nu}$ the coordinates are degenerate. We assume a domain where ${x_\mu\neq x_\nu}$ for ${\mu\neq\nu}$. In such a domain we can always order and rescale the coordinates in such a way that
\begin{equation}\label{coorord}
    x_\mu + x_\nu > 0 \quad\text{and} \quad x_\mu - x_\nu > 0 \quad \text{for}\;\mu<\nu\;.
\end{equation}
With this convention and assuming Euclidian signature we have
\begin{equation}\label{eq:module}
    U_\mu = (-1)^{N{-}\mu} |U_\mu|\;,\quad X_\mu=(-1)^{N{-}\mu} |X_\mu|\;.
\end{equation}

The metric can be written in the diagonal form
\begin{equation}\label{diagmetric}
{g}= \sum_{\mu}\,
    \Bigl(\,\ef^\mu \otimes \ef^\mu
    + \ef^{\hat \mu} \otimes \ef^{\hat\mu}\,\Bigr) + \eps\, \ef^0 \otimes \ef^0
  \;,
\end{equation}
where we have introduced an orthonormal covector frame $\ef^a={\{\ef^\mu,\ef^{\hat\mu}, \ef^0\}}$,
\begin{equation}  \label{formframe}
\begin{gathered}
\ef^\mu = \frac{d x_{\mu}}{\sqrt{Q_\mu}}\;,\quad
  \ef^{\hat\mu} = \sqrt{Q_\mu} \sum_{j=0}^{N-1}\A{j}_{\mu}d\psi_j   \;, \\
\ef^0 = \sqrt{S} \sum_j \A{j}d\psi_j  \, .
\end{gathered}
\end{equation}
The dual vector frame $\ef_a={\{\ef_\mu,\ef_{\hat\mu}, \ef_0\}}$ is
\begin{gather}
 \ef_\mu = \sqrt{Q_\mu}\cv{x_{\mu}}\;,\quad
 \ef_{\hat\mu} = \sqrt{Q_\mu}\sum_{j}\frac{(-x_\mu^2)^{N{-}1{-}j}}{X_\mu}\cv{\psi_j} \; , \nn\\
 \ef_0 = \frac{1}{\sqrt{S} \A{N}} \cv{\psi_N} \, . \label{vectfr}
\end{gather}
Note that $\ef^0$ and $\ef_0$ are defined only in an odd dimension.
The Ricci tensor is also diagonal in this frame and is given by \cite{HamamotoEtal:2007}
\ba\label{Ric}
    \Ric &=& \sum_{\mu} r_\mu
    \Bigl(\,\ef^\mu \otimes \ef^\mu
    + \ef^{\hat \mu} \otimes \ef^{\hat\mu}\,\Bigr) + \eps r_0 \ef^0 \otimes \ef^0  \;,\nonumber\\
r_\mu &=& - \frac1{2x_\mu}\biggl[\sum_{\nu}
\frac{x_\nu^2\bigl(x_\nu^{-1} \hat{X}_\nu\bigr){}_{,\nu}}{U_\nu} + \eps \sum_\nu \frac{\hat{X}_\nu}{U_\nu} \biggr]_{\!,\mu}\; , \nn \\
r_0 &=& - \sum_\rho \frac{1}{x_\rho} \left( \sum_\sigma \frac{\hat{X}_\sigma}{U_\sigma} \right)_{\! \!\!, \rho} \,,\ \ \!\hat{X}_\mu = X_\mu - \frac{\eps c}{x_\mu^2} \, \,.
\ea
The PKY tensor of the canonical metric reads
\begin{equation}\label{PCKY}
    h = \sum_{\mu=1}^N x^\mu\, \ef^\mu\wedge\ef^{\hat\mu}\;.
\end{equation}
It generates a tower of even rank CCKY forms $\CCKY{j}$,  and even or odd rank (in even or odd number of spacetime dimensions) KY tensors $\KY{j}$,
\ba
\CCKY{j}&=&\frac1{j!}\, h\wedge\dots\wedge h\;,\label{CCKY}\\
\KY{j}&=&*\CCKY{j}=(-1)^j\,\LCT\,\CCKY{j}\,,\label{fj}
\ea
as well as the 2nd-rank Killing tensors ($j=0,\dots, N-1$)
\begin{equation}\label{KT}
\KT{j}=\sum_{\mu}\A{j}_\mu
\Bigl(\ef^\mu \otimes  \ef^\mu\! +\!
\ef^{\hat \mu}\otimes\ef^{\hat \mu}\Bigr) + \eps \A{j} \ef^0 \otimes  \ef^0   \;.
\end{equation}
The primary Killing vector $\PKV$, \eqref{primary}, coincides with $\KV{0}$ given by \eqref{KV}, ${\PKV}=\KV{0} =\cv{\psi_0}$, and can be
explicitly written as
\be \label{PKV}
  \PKV = \sum_\mu \sqrt{Q_\mu} \ef_{\hat{\mu}} + \eps \sqrt{S} \ef_0\,.
\ee
It satisfies an important relation
\begin{equation}\label{nablaCCKY}
    -\frac1{n{-}2j{+}1}\,\delta\CCKY{j} = \PKV^\flat \wedge\CCKY{j{-}1}\;.
\end{equation}
The secondary Killing vectors can also be generated from the PKY tensor, e.g., \cite{KrtousEtal:2007jhep}, and are explicitly given by
\be\label{KVs}
\KV{k} = \sum_\mu \sqrt{Q_\mu} \A{k}_\mu \ef_{\hat{\mu}} + \eps \A{k}\sqrt{S} \ef_0\,.
\ee
In odd dimensions we have
\be\label{cood}
\KVf{N}=(\partial_{\psi_N})^\flat=\sqrt{-c}*\CCKY{N}=\sqrt{-c}\,\KY{N}\,.
\ee
All the Killing vectors mutually Lie-bracket-commute, as well as preserve all the Killing--Yano and Killing tensors constructed from $h$,
e.g.,
\be\label{property}
{\cal L}_{\KV{k}} \KV{j}=0\,,\quad {\cal L}_{\KV{k}}\CCKY{j}=0\,.
\ee
We conclude with a few identities that will be used in Sec.~\ref{sec4}.
The following two relations have been proved in \cite{CarigliaEtal:2011a}:
\begin{gather}
  \bigl[ \CCKY{k},\,\CCKY{l} \bigr] = 0 \,,\label{hh=0}\\
  \bigl[ e^{(a}\wedge\CCKY{k},\,e^{b)}\wedge \CCKY{l}\bigr]=0\,.\label{seheh=0}
\end{gather}
The latter identity implies
\begin{equation}
  \bigl[ \omega\wedge\CCKY{k},\,\omega\wedge \CCKY{l}\bigr]=0\label{ohoh=0}
\end{equation}
for any 1-form ${\omega}$. In Appendix~\ref{appA} we also prove
\begin{equation}
  \nabla_{\![a}\PKV_{b]}\;\bigl[ e^{[a}\wedge\CCKY{k},\,e^{b]}\wedge \CCKY{l}\bigr]_+ = 0\;.\label{Faeheh+=0}
\end{equation}
In all these identities, ${[\;,\,]}$ and ${[\;,\,]_+}$ denote Clifford commutator and anti-commutator, respectively.
Other properties of canonical metrics are gathered, for example, in recent reviews \cite{Yasui:2011,Cariglia:2012}.

\subsection{Test electromagnetic field} \label{subs:sec32}
We shall now equip the canonical spacetime with the test electromagnetic field $A$ given by a linear combination of background isometries ${\KVf{k}}$,
\be\label{A}
A=\sum_k q_k\, \KVf{k}\,,
\ee
with constants ${q_k}$ characterizing the field strength.
In order this field to be a solution of the vacuum Maxwell equations,
we have to impose a condition $\Ric_{ab}A^b=0$, cf. Eqs. \eqref{eq:Maxwell_curved} and \eqref{eq:Killing_curved}. Taking into account the diagonal form \eqref{Ric} of the Ricci tensor, the explicit form \eqref{KVs} of the Killing vectors and a functional independency of the functions ${\A{j}_\mu}$, we find that spacetime must be a vacuum {\em Kerr-NUT spacetime} \cite{ChenEtal:2006cqg},
${\Ric=0}$, with the metric functions ${X_\mu}$ in the form\footnote{%
We impose the condition of vanishing electric current to guarantee a realistic matter content of the spacetime. However, the commutativity of the symmetry operators and the separability of the Dirac equation discussed below are independent of this requirement. For example, we can relax the vacuum condition for the metric and consider a nontrivial cosmological constant. In such a case, the spacetime must be filled with a charged test fluid, with electric current aligned with the vector potential, i.e., with the chosen isometry of the spacetime.}
\begin{equation}\label{BHXs}
  X_\mu = \sum_{k=\eps}^{N{-}1}\, c_{k}\, x_\mu^{2k} - 2 b_\mu\, x_\mu^{1-\eps} + \frac{\eps c}{x_\mu^2} \;
\end{equation}
It describes a general rotating black hole with spherical horizon topology.
Parameters $c_k, c$ and $b_\mu$ are related to $(N+\eps)$ independent angular momenta,
a mass parameter, and NUT charges.

After fixing the metric, let us now study what restrictions the test electromagnetic field $A$ has to satisfy in oder it to be compatible with the existence of $n$ first order operators commuting with the charged Dirac operator. Applying the results of Sec.~\ref{sec2}, in any dimension the candidates for such operators are generated by $(N+\eps)$ Killing forms $\KVf{k}$ and $N$ CCKY tensors $\CCKY{j}$.\footnote{Odd KY tensors $\KY{j}$ in odd dimensions are dual to even CCKY tensors $\CCKY{j}$. Hence due to the fact that both, the operators \eqref{eq:M_to_K} and the anomaly condition \eqref{cond}, are Hodge dual, without loss of generality we may focus on operators generated from CCKY tensors $\CCKY{j}$ only.}
The corresponding operators ${K_{\KVf{k}}}$, cf.~\eqref{Kk}, and ${M_{\CCKY{j}}}$, cf.~\eqref{MdefA}, commute with ${\cal D}$ only if the anomalous conditions \eqref{Kk} and \eqref{cond} are satisfied.

Since all the isometries $\KV{k}$ of the Kerr-NUT spacetime mutually commute [first property \eqref{property}], the anomaly condition
\be
{\cal L}_{\KV{k}}A=0
\ee
is satisfied and the operators \eqref{Kk} commute with ${\cal D}$ for any $A$ given by \eqref{A}.

However, for the operators \eqref{MdefA} the situation is more complicated.
Let us start from the second property \eqref{property} and re-express the Lie derivative of $h$ along $A$ as follows:
\ba
0&=&{\cal L}_{A^\sharp} h=d(A \cwedge{1}h)\nonumber\\
&=& - dA \cwedge{1}h+\nabla_{\!a} A \wedge X^a\hook h+\nabla_{\!A^\sharp} h\nonumber\\
&=&-\frac{1}{2}dA \cwedge{1}h+A \wedge \PKV^\flat\,,
\ea
where we have used the fact that $A$ is a Killing form and $h$ a PKY tensor.
Obviously, only for $A$ proportional to the primary Killing form $\PKV^\flat$, the latter term vanishes and the anomaly condition ${\displaystyle dA \cwedge{_1}h=0}$ is satisfied. The same remains true for higher rank CCKY tensors
$\CCKY{j}$. Namely, the anomaly condition
\be
dA \cwedge{1}\CCKY{j}=0
\ee
is satisfied only for such $A$ given by \eqref{A} which is proportional to the primary Killing form,
\begin{equation}\label{Apropexi}
  A = e\,\PKV^{\flat}\;,
\end{equation}
i.e., ${q_0=e}$ and ${q_k=0}$ otherwise. Hence, from the `mathematical point of view' the test electromagnetic field determined from the primary Killing vector is fundamentally  different from the test electromagnetic field determined by the secondary Killing fields; only when ${A=e\, \PKV^\flat}$, all $\dm$ first order operators \eqref{Kk} and \eqref{MdefA} commute with the charged Dirac operator ${\cal D}$.

Physically, the test electromagnetic field \eqref{Apropexi} given by the primary Killing vector describes a {\em weakly charged} Kerr-NUT spacetime, whereas the test field associated with the secondary Killing vectors corresponds to a black hole immersed in a {\em magnetic field}.

\subsection{Charged scalar particle}
It is well known that in the canonical spacetime in the absence of a test electromagnetic field, the Hamilton--Jacobi equation as well as the Klein--Gordon and Dirac field equations allow separation of variables \cite{FrolovEtal:2007, OotaYasui:2008}.

The motion of a scalar charged particle in the weakly charged or magnetized spacetimes
has been studied more recently \cite{FrolovKrtous:2011}. It was observed that in the weakly charged Kerr-NUT spacetime the motion of a charged scalar particle maintains its complete integrability, while the corresponding charged Hamilton--Jacobi equation separates. This integrability can be characterized by the existence of a nontrivial
Lax tensor discussed in \cite{CarigliaEtal:2012}.
The separability property remains true also for a charged scalar field in the vicinity of the weakly charged black hole. Such separability is underlain by the existence of a complete set of mutually commuting symmetry operators generated from Killing vectors $\KV{k}$ and Killing tensors $\KT{j}$, \eqref{KT},
\ba
{\cal L}_{(k)}&=&-i\KV{k}^{a}\nabla_a\,,\nonumber\\
{\cal K}_{(j)}&=&-\left[\nabla_a-iA_a\right]\KT{j}{}^{ab}\left[\nabla_b-iA_b\right]\,,
\ea
with $A$ given by \eqref{Apropexi}.

On the other hand, it has been observed in \cite{FrolovKrtous:2011} that similar operators do not commute when
instead of the primary Killing field  $\PKV^\flat$, $A$ is associated with the secondary Killing fields, i.e., in the immersed magnetic field. In this case neither the charged Hamilton--Jacobi, nor the charged scalar field equations separate in canonical coordinates.

The aim of this paper is to demonstrate that a similar conclusion remains valid for a charged spin
$\frac{1}{2}$ field. We already know that in the weakly charged case there exist $n-1$ operators commuting with the charged Dirac operator ${\cal D}$. In the next section we demonstrate that all these operators mutually commute and hence together with ${\cal D}$ form a complete set of symmetry operators. In Sec.~\ref{sec6} the explicit separation of the charged Dirac equation in the chosen representation will be demonstrated.

\section{Complete set of Dirac symmetry operators}\label{sec4}

\subsection{Complete set of operators}\label{ssc:csop}
The canonical spacetime \eqref{metric} in the absence of a test electromagnetic field allows
a complete separation of variables for the Dirac field equation \cite{OotaYasui:2008}. It was shown in \cite{CarigliaEtal:2011a, CarigliaEtal:2011b} that
such a separability can be characterized by the existence of a complete set of mutually commuting first order symmetry operators.  These operators are determined by the tower of symmetries built from the PKY tensor ${h}$. Namely, they are given by $({N}+\eps)$ KY 1-forms $\KVf{k}$, \eqref{KV}, and ${N}$ CCKY forms ${\CCKY{j}}$, \eqref{CCKY}. In the exterior algebra notation they read:
\begin{align}
  K^{(0)}_k &= K^{(0)}_{\KVf{k}} = X^a\hook \KVf{k}\nabla_{a} + \frac{1}{4}d\KVf{k} \, ,
    \label{unchargedops}\\
  M_j^{(0)} &= M^{(0)}_{\CCKY{j}} = e^a\wedge\CCKY{j}\nabla_{\!a} + \frac12 (n{-}2j)\,\PKV\wedge \CCKY{j-1}
    \notag
\end{align}
In the expression for ${M_j^{(0)}}$ we have employed \eqref{nablaCCKY} to simplify the second term.
Note also that the operator $M_0^{(0)}$ coincides with the Dirac operator, $M_0=D$.

In the case of weakly charged Kerr-NUT spacetimes discussed in the previous section, the operators \eqref{unchargedops} can be upgraded, to give a complete set of mutually commuting operators of the charged Dirac operator ${\cal D}$, \eqref{chargedD}.  Following the results \eqref{Kk} and \eqref{MdefA} from Sec.~\ref{sec2}, together with the discussion in Sec.~\ref{subs:sec32}, the operators commuting with the charged Dirac operator are
\begin{align}
  K_k &= K_{\KVf{k}} = K^{(0)}_k \, , \label{chargedKops}\\
  M_j &= M_{\CCKY{j}} = M^{(0)}_j-ie\,\PKV\wedge \CCKY{j}\,,\label{chargedMops}
\end{align}
with $k = 0, \dots, N-1+\eps$  and $j = 0, \dots, N-1$.
We shall now show that, similarly to the uncharged case, the operators \eqref{chargedKops} and \eqref{chargedMops} mutually commute.


\subsection{Mutual commutativity}\label{ssc:comop}
 It was shown in \cite{CarigliaEtal:2011a} that the commutator with the operator ${K_\kappa}$ is equivalent to a Lie derivative along a Killing vector ${\kappa^\sharp}$. The mutual commutativity of \eqref{chargedKops} thus follows from the Lie-commutativity of the Killing vectors, first property \eqref{property}, while the commutativity of operators \eqref{chargedKops} with operators \eqref{chargedMops} follows from the second property \eqref{property}.

It remains to prove that also all the operators \eqref{chargedMops} mutually commute. The commutator can be split into the following six terms:
\begin{equation}\label{Mcom}
\begin{split}
  &\bigl[M_{k},\,M_{l}\bigr]
    =\bigl[M^{(0)}_{k},\,M^{(0)}_{l}\bigr]\\
  &\quad-2ie\,\PKV_a\,\bigl[e^{(a}\wedge\CCKY{k},\,e^{b)}\wedge\CCKY{l}\bigr]\nabla_{\!b}\\
  &\quad+2ie\, (\nabla_{\![a}\PKV_{b]}) \bigl[e^{[a}\wedge\CCKY{k},\,e^{b]}\wedge\CCKY{l}\bigr]_+\\
  &\quad-ie\,\frac{\dm{-}2k}2\bigl[\PKV\wedge\CCKY{k-1},\,\PKV\wedge\CCKY{l}\bigr]\\
  &\quad-ie\,\frac{\dm{-}2l}2\bigl[\PKV\wedge\CCKY{k},\,\PKV\wedge\CCKY{l-1}\bigr]\\
  &\quad-e^2\,\bigl[\PKV\wedge\CCKY{k},\,\PKV\wedge\CCKY{l}\bigr]\,,
\end{split}
\end{equation}
where we have used ${\PKV\wedge\nabla_{\!a}\CCKY{k} = 0}$, cf.~\eqref{nablaCCKY}. Employing the commutativity of uncharged operators ${M^{(0)}_k}$, and the identities \eqref{seheh=0}--\eqref{Faeheh+=0}, we find that all the terms independently vanish and  hence the operators \eqref{chargedMops} commute.
[An alternative proof using the explicit representation of operators $M_j$ is gathered in App.~\ref{appB}.]

\section{Explicit representation}\label{sec5}
In this section we find an explicit representation of the action of operators \eqref{chargedKops} and \eqref{chargedMops} on the Dirac bundle in which the charged Dirac equation separates.
As customary, we shall omit the explicit symbol $\gamma_*$, defined by \eqref{cleaiso}, whenever it is clear from context that operators are being considered instead of forms. So, for example, we write $K_j$ for $\gamma_*K_j$ and $M_j$ for $\gamma_*M_j$.

\subsection{Spinors and $\gamma$ matrices}
Following \cite{CarigliaEtal:2011b}, we represent the fiber of the Dirac bundle as a tensor product of ${N}$ 2-dimensional spaces ${\fiber{S}}$, ${\fiber{D}M=\fiber{S}^N M}$. We use Greek letters ${\epsilon=\pm,\varsigma=\pm,\dots}$ for tensor indices in these 2-dimensional spaces and write
a generic 2-dimensional spinor as $\chi = \chi^+ \tht_+ + \chi^- \tht_- = \chi^\epsilon \tht_\epsilon$. Here $\tht_+$ and $\tht_-$ represent a frame in the 2-dimensional spinor space~${\fiber{S}}$ and the components of the spinor are two complex numbers ${\left( \begin{smallmatrix} \chi^+ \\ \chi^- \end{smallmatrix} \right)}$.
The Dirac spinor ${\psi\in\fiber{D} M}$ is then written as
\be\label{spinor}
\psi = \psi^{\epsilon_1\dots\epsilon_N} \tht_{\epsilon_1\dots\epsilon_N}
\ee
with ${2^N}$ components ${\psi^{\epsilon_1\dots\epsilon_N}}$ and the Dirac bundle frame
${\tht_E}$ is given by
\begin{equation}\label{Thetafr}
    \tht_E = \tht_{\epsilon_1\dots\epsilon_N}
      = \tht_{\epsilon_1}\otimes\dots\otimes\tht_{\epsilon_N}\;.
\end{equation}

The gamma matrices are constructed as various tensor products of Pauli matrices. Let ${\one}$, ${\iota}$, ${\sigma}$, and ${\hat\sigma}$ be the unit and Pauli operators on ${\fiber{S}}$, respectively. In components, their action reads
\begin{equation}\label{sigmamatr}
\begin{gathered}
    (\one\,\chi)^\epsilon = \chi^\epsilon\;,\quad
    (\iota\,\chi)^\epsilon = \epsilon\,\chi^\epsilon\;,\\
    (\sigma\,\chi)^\epsilon = \chi^{-\epsilon}\;,\quad
    (\hat\sigma\,\chi)^\epsilon = -i\epsilon\,\chi^{-\epsilon}\;.
\end{gathered}
\end{equation}
In matrix form they are written as:
\begin{equation}\label{sigmamatrcomp}
\begin{gathered}
    \one^\epsilon{}_\varsigma \equiv
    \left(\begin{array}{cc}
        1 & 0 \\
        0 & 1 \\
      \end{array}\right)
    \;,\quad
    \iota^\epsilon{}_\varsigma \equiv
    \left(\begin{array}{cc}
        1 & 0 \\
        0 & -1 \\
      \end{array}\right)
    \;,\\
    \sigma^\epsilon{}_\varsigma \equiv
    \left(\begin{array}{cc}
        0 & 1 \\
        1 & 0 \\
      \end{array}\right)
    \;,\quad
    \hat\sigma{}^\epsilon{}_\varsigma \equiv
    \left(\begin{array}{cc}
        0 & -i \\
        i & 0 \\
      \end{array}\right)
    \;.
\end{gathered}
\end{equation}
For any linear operator ${\alpha\in\fiber{S}^1_1M}$ we denote by ${\alpha_{\lst{\mu}}\in\fiber{D}^1_1M}$ a linear operator on the Dirac bundle
\begin{equation}\label{alphaDB}
    \alpha_{\lst{\mu}} \equiv \one\otimes\dots\otimes\one\otimes\alpha\otimes\one\otimes\dots\otimes\one
\end{equation}
with ${\alpha}$ on the ${\mu}$-th place in the tensor product. Similarly, for mutually different indices ${\mu_1,\dots,\mu_j}$ we define
\begin{equation}\label{multialphaDB}
    \alpha_{\lst{\mu_1\dots\mu_j}} \equiv \alpha_{\lst{\mu_1}}\otimes\dots\otimes\alpha_{\lst{\mu_j}}\;.
\end{equation}

Using these preliminaries, the abstract gamma matrices with respect to the frame $\ef_a=\{{\ef_\mu,\,\ef_{\hat\mu}}, \ef_0\}$ can be written as
\begin{equation}\label{gammamatrmuhatmu}
\begin{gathered}
    \gamma^\mu = \iota_{\lst{1\dots\mu{-}1}}\sigma_{\lst{\mu}}\;,\quad
    \gamma^{\hat\mu} = \iota_{\lst{1\dots\mu{-}1}}\hat\sigma_{\lst{\mu}}\, , \\
    \gamma^{0} = \iota_{\lst{1 \dots N}} \, ,
\end{gathered}
\end{equation}
with $\gamma^{0}$ defined only in odd dimension.
In components, the action of these matrices on a spinor \eqref{spinor} reads
\begin{equation}\label{gammaaction}
\begin{aligned}
    (\gamma^\mu\psi)^{\epsilon_1\dots\epsilon_N} &=
     \Bigl(\prod_{\nu =1}^{\mu-1} \epsilon_\nu\Bigr)\,\psi^{\epsilon_1\dots(-\epsilon_\mu)\dots\epsilon_N}\;,\\
    (\gamma^{\hat\mu}\psi)^{\epsilon_1\dots\epsilon_N} &=
     -i\epsilon_\mu\,\Bigl(\prod_{\nu=1}^{\mu-1} \epsilon_\nu\Bigr)\psi^{\epsilon_1\dots(-\epsilon_\mu)\dots\epsilon_N}\;, \\
    (\gamma^{0}\psi)^{\epsilon_1\dots\epsilon_N} &=
     \Bigl(\prod_{\nu =1}^{N} \epsilon_\nu\Bigr)\,\psi^{\epsilon_1\dots \epsilon_N}\;.
\end{aligned}
\end{equation}
Note the relations
\be\label{relG}
\gamma^{\hat \mu}=-i\iota_{\lst{\mu}}\gamma^\mu\,,\quad  \gamma^{\mu \hat \mu}=i \iota_{\lst{\mu}}\,
\ee
and the fact that in odd dimensions we have
\be\label{gammaz}
\gamma_*(\LCT)=\gamma^{\mu_1 \dots \mu_N\hat \mu_1\dots \hat \mu_N\,0}=i^N\,.
\ee

\subsection{Symmetry operators}
Symmetry operators \eqref{chargedKops} corresponding to Killing vectors are, in general, equivalent to the Lie derivative lifted from the tangent bundle to the Clifford or Dirac bundles. Thanks to \eqref{KV} we thus write
\begin{equation}\label{opKexpl}
    K_{k} = 
 \cd{\psi_k}\;,
\end{equation}
where ${\cd{\psi_k}}$ is a partial derivative along ${\psi_k}$, acting just on the components of the spinor with respect to frame $\tht_E$, i.e.,  $K_{\!k}\chi={\cd{\psi_k}{\chi}=\frac{\partial\chi^E}{\partial \psi_k}\tht_E}$.

To lift the operators ${M_j}$, \eqref{chargedMops}, to the Dirac bundle we employ \eqref{cleaiso}. Using the results in \cite{CarigliaEtal:2011b} we have the following important formula valid for any 1-form ${\alpha}$:
\begin{equation}\label{ClfaCCKY}
  \gamma_*\bigl(\alpha\!\wedge\!\CCKY{j}\bigr) = i^j \!\sum_\mu \Bo{j}_\mu
       \bigl(\alpha_\mu \gamma^\mu\! +\!\alpha_{\hat\mu}\gamma^{\hat\mu} \bigr)\!+\!\eps i^j \Bo{j} \alpha_{0} \gamma^{0} \;.
\end{equation}
Here, $\Bo{k}_{\mu}$ and $\Bo{k}$ represent a `spinorial analogue' of functions ${\A{j}_\mu}$ and ${\A{j}}$,  \eqref{eq:A_def}, and read
\ba
\Bo{k}_{\mu}&=&\hspace{-3mm}
    \sum\limits_{\substack{\nu_1,\dots,\nu_k\\\nu_1<\dots<\nu_k,\;\nu_i\ne\mu}}\!\!\!\!\!
    \iota_{\lst{\nu_1}}x_{\nu_1}\cdots\ \iota_{\lst{\nu_k}}x_{\nu_k} \;,\nonumber\label{B_mu_def}\\
  \Bo{k}&=& \hspace{-1mm}\sum\limits_{\substack{\nu_1,\dots,\nu_k\\\nu_1<\dots<\nu_k}}\!\!\!\!\!
    \iota_{\lst{\nu_1}}x_{\nu_1}\cdots\ \iota_{\lst{\nu_k}}x_{\nu_k} \;.  \label{Bdef}
\ea
In particular, we have
\begin{equation}\label{0ordcoef}
\begin{split}
&\gamma_*\Bigl(\PKV^\flat\wedge\CCKY{j}\Bigr)\\
&\quad=-i^{j+1} \Bigl(\,
    \sum_\mu\!\sqrt{Q_\mu}\,\Bo{j}_\mu\iota_{\lst{\mu}}\gamma^{\mu}
+  i \eps \sqrt{S} \Bo{j} \gamma^0 \Bigr)\, .
\end{split}\raisetag{10ex}
\end{equation}
Next, one has to use the expression for a spinorial derivative
\begin{equation}\label{covdfr}
    \nabla_a = \eth_a + \frac14\omega_{abc}\gamma^b\gamma^c\;,
\end{equation}
where  ${\eth_a}$ is a derivative acting only on components of the spinor and the spin connection coefficients are gathered in appendix of \cite{CarigliaEtal:2011b}. The resulting formula for $M_j$ reads
\begin{equation}\label{Mopexpl}
\begin{split}
M_j
&= i^j \!\sum_\mu\! \sqrt{Q_\mu} \Bo{j}_\mu\Biggl(
    \cd{x_\mu}\!+\!\frac{X_\mu'}{4X_\mu} \!+\!  \frac12 \sum_{\substack{\nu\\\nu\neq\mu}} \frac1{x_\mu\!-\!\iota_{\lst{\mu\nu}}x_\nu}  \\
-& \frac{i\iota_{\lst{\mu}}}{X_\mu}\!\sum_k (-x_\mu^2)^{N{-}1{-}k}\cd{\psi_k}
    \!+\!\frac{\eps}{2x_\mu} \!-\! e\iota_{\lst{\mu}}\Biggr)\gamma^\mu   \\
+& \eps\, i^{j+1} \frac{\sqrt{S}}{2} \Biggl[
   \Bo{j-\!1}\!-\!\Bo{j}\Bigl(\frac{2}{ic}\cd{\psi_N}\!+\!\!\sum_\mu\! \frac{1}{\iota_{\lst{\mu}} x_\mu} \!+\! 2e\Bigr)  \Biggr] \gamma^0 \, .
\end{split}\raisetag{13.5ex}
\end{equation}

\section{Separability of the charged Dirac equation}\label{sec6}
Let us formulate the main result: the commuting symmetry operators ${K_k}$ and ${M_j}$ have common spinorial eigenfunctions ${\psi}$
\begin{align}
    K_k \psi &= i\,\psc{k}\psi\;,\label{eigenfcK}\\
    M_j \psi &= \chc{j}\psi\;,\label{eigenfcM}
\end{align}
which can be found in the tensorial R-separated form
\begin{equation}\label{tensRsep}
    \psi = R\,\exp\bigl({\textstyle i\sum_k\psc{k}\psi_{k}}\bigr)\,
           \bigotimes_\nu \chi_\nu\;.
\end{equation}
Here $\left\{\chi_\nu \right\}$ is an $N$-tuple of 2-dimensional spinors, where each spinor $\chi_\nu$ depends only on the variable ${x_\nu}$, $\chi_\nu=\chi_\nu(x_\nu)$, and ${R}$ is the (Clifford bundle)-valued prefactor
\begin{equation}\label{Phidef}
  R = \prod_{\substack{\kappa,\lambda\\\kappa<\lambda}}
    \Bigl(x_\kappa+\iota_{\lst{\kappa\lambda}}x_\lambda\Bigr)^{-\frac12}\;.
\end{equation}
Eqs. \eqref{eigenfcK} and \eqref{eigenfcM} are satisfied if and only if the spinors $\chi_\nu$ satisfy the ordinary differential equations \eqref{chieq} below, which generalize the conditions found in \cite{OotaYasui:2008} and \cite{CarigliaEtal:2011b} for the case $e = 0$.

%

The spinor ${\psi}$ given by \eqref{tensRsep} clearly satisfies \eqref{eigenfcK}. To show \eqref{eigenfcM}, we need to calculate $M_j\psi$, with $M_j$ given by \eqref{Mopexpl}.
We have
\begin{equation}\label{Mjpsi1}
\begin{split}
&M_j \psi=i^j \exp\bigl({\textstyle i\sum_k\psc{k}\psi_{k}}\bigr)
  \Biggl[\!\sum_\mu\! \sqrt{Q_\mu} \Bo{j}_\mu\times  \\
& \times\Biggl(
    \cd{x_\mu}{+}\frac{X_\mu'}{4X_\mu}
    {+}\frac12 \sum_{\substack{\nu\\\nu\neq\mu}} \frac1{x_\mu{-}\iota_{\lst{\mu\nu}}x_\nu}
    {+}\frac{\tilde \Psi_\mu}{X_\mu}\iota_{\lst{\mu}}
    {+}\frac{\eps}{2x_\mu}\Biggr)\gamma^\mu  \\
&+ \eps\, \frac{i\sqrt{S}}{2}
   \Biggl(\! \Bo{j\!-\!1}\!\!-\!\Bo{j}\Bigl(\frac{2 \Psi_N}{c}
   \!+\!\sum_\mu\! \frac{\iota_{\lst{\mu}}}{x_\mu} \!+\! 2e\Bigr) \Biggr) \gamma^0 \!\Biggr]\,R\bigotimes_\nu  \chi_\nu\;,\\
\end{split}\raisetag{22.7ex}
\end{equation}
where we have performed the derivative with respect to angles ${\psi_k}$ and introduced the functions ${\psf{\mu}}$ of one variable ${x_\mu}$ given by
\begin{equation}\label{psfdef}
    \psf{\mu} = \sum_k \psc{k} (-x_\mu^2)^{N{-}1{-}k} - eX_\mu\;.
\end{equation}
Let us now concentrate on the derivatives of the prefactor ${R}$. Using the following relations
proved in \cite{CarigliaEtal:2011b}:
\ba
R^{-1}\gamma^\mu R &=&  \frac{\sqrt{|U_\mu|}}{\Vo_{\mu}} \left( - \iota_{\lst{\mu}}\right)^{N-\mu} \sigma_{\lst{\mu}} \, ,\label{eq:gamma_mu_phi}\\
\gamma^\mu \frac{\eth R}{\partial{x_\mu}} &=&
\Bigl( - \frac12 \sum_{\substack{\nu\\\nu\neq\mu}} \frac1{x_\mu\!{-}\iota_{\lst{\mu\nu}}x_\nu} \Bigr) \gamma^\mu R \,,\label{derPhi}
\ea
where
\be
\Vo_{\mu}=\prod\limits_{\substack{\nu\\\nu\ne\mu}} (\iota_{\lst{\nu}} x_{\nu}-\iota_{\lst{\mu}}x_{\mu})\;,\label{Vdef}
\ee
we can bring the operator $R$ to the front, to get
\begin{equation}\label{Mjpsi2}
\begin{split}
&M_j\psi =i^j \exp\bigl({\textstyle i\sum_k\!\psc{k}\psi_{k}}\bigr)\,R\\
&\times\!\Biggl[\sum_\mu \!\frac{\sqrt{|X_\mu|}}{\Vo_{\mu}}
  \bigl(-\! \iota_{\lst{\mu}} \bigr)^{\!N\!-\!\mu}\Bo{j}_\mu\\
&\qquad\quad\quad\times\Bigl( \cd{x_\mu}+\frac{X_\mu'}{4X_\mu}
  +\frac{\tilde \Psi_\mu}{X_\mu}\iota_{\lst{\mu}}
  +\frac{\eps}{2x_\mu}\Bigr)\,\sigma_{\lst{\mu}} \\
&\quad+\! \eps\, \frac{i\sqrt{S}}{2} \Bigl(\! \Bo{j\!-\!1}
  {-}\Bo{j}\Bigl(\frac{2 \Psi_N}{c}{+}\sum_\mu\! \frac{\iota_{\lst{\mu}}}{x_\mu} \!+\! 2e\Bigr)
  \Bigr)\, \gamma^0 \Biggr]
  \bigotimes_\nu \chi_\nu\,,
\end{split}\raisetag{16ex}
\end{equation}
which is to be compared with
\be\label{msi3}
\chc{j}\psi= \chc{j}\exp\bigl({\textstyle i\sum_k\psc{k}\psi_{k}}\bigr)R
           \bigotimes_\nu \chi_\nu \,.
\ee
Following \cite{CarigliaEtal:2011b} we further introduce the following functions $\chf{\nu}$ of a single variable ${x_\nu}$:
\be  \label{eq:chf_mu_definition}
\chf{\nu} =  \sum_{j} (-i)^j \chc{j} \left( -\iota_{\lst{\nu}} x_\nu \right)^{N-1-j} \, .
\ee
Note that in odd dimensions the constant $\chc{N}$, defined by $M_N\psi=\chc{N}\psi$, is not independent.
In fact, using the relation between operators $K$ and $L$, the Hodge duality \eqref{eq:M_to_K} together with \eqref{gammaz}, and Eqs. \eqref{eigenfcK}, \eqref{eigenfcM}, we have 
\be\label{chiN}
\chc{N}=\frac{i^{N+1}}{\sqrt{-c}}\bigl(\Psi_N+ce\bigr) \, .
\ee

We are now ready to derive the differential equations for $\chi_\nu$ so that \eqref{eigenfcM} are satisfied.
We can cancel the common $\exp\bigl({\textstyle i\sum_k\psc{k}\psi_{k}}\bigr) R$ prefactor in \eqref{Mjpsi2} and
\eqref{msi3}  (in the coordinate domain we are using the operator $R$ is never zero on any spinor), multiply both equations by $(-i)^j \left( -\iota_{\lst{\nu}} x_\nu \right)^{N-1-j}$ and sum over $j$ to obtain
\begin{equation}\label{predpos}
\begin{split}
&\chf{\nu} \bigotimes_\kappa  \chi_\kappa=\\
&=\Biggl[\!\sqrt{|X_\nu|}\bigl(-\! \iota_{\lst{\nu}} \bigr)^{\!N\!-\!\nu}
  \Bigl(\cd{x_\nu}{+}\frac{X_\nu'}{4X_\nu}
  {+}\frac{\tilde \Psi_\nu}{X_\nu}\iota_{\lst{\nu}}
  {+}\frac{\eps}{2x_\nu}\Bigr)\sigma_{\lst{\nu}}\\
&\mspace{100mu}- \eps\, \frac{i\sqrt{S}}{2x_\nu^2} \Bo{N}\gamma^0 \Biggr]\,
  \bigotimes_\kappa  \chi_\kappa\,\,.
\end{split}\raisetag{5ex}
\end{equation}
Here we have used the `completeness relations'
 \begin{equation}\label{BVrel}
\begin{gathered}
\sum_{\mu} \frac{\Bo{i}_\mu}{\Vo_\mu}\,
{\bigl(-\iota_{\lst{\mu}}x_\mu\bigr)^{N{-}1{-}j}} =
  \delta^i_j\;,\\
\sum_{j}\frac{\Bo{j}_\mu}{\Vo_\nu}\, {\bigl(-\iota_{\lst{\nu}}x_\nu\bigr)^{N{-}1{-}j}} = \delta^\nu_\mu\;,
\end{gathered}
\end{equation}
together with the following identities:
\ba
&&\sum_{j=0}^N \Bo{j} \left( -\iota_{\lst{\mu}} x_\mu \right)^{N-1-j} = 0 \, ,\label{eq:sum_Boj}\\
&&\sum_{j=0}^N \Bo{j-1} \left( -\iota_{\lst{\mu}} x_\mu \right)^{N-1-j} = - \frac{\Bo{N}}{x_\mu^2} \, . \label{eq:sum_Boj-1}
\ea
Using further the formula
\be\label{Sid}
 \gamma^0 \sqrt{S} = \frac{\sqrt{-c}}{\Bo{N}} \, ,
\ee
we can rewrite Eq. \eqref{predpos} as
\begin{equation}\label{pos}
\begin{split}
&\Biggl[\!\sqrt{|X_\nu|}\bigl(-\! \iota_{\lst{\nu}} \bigr)^{\!N\!-\!\nu}
\Bigl(
    \cd{x_\nu}{+}\frac{X_\nu'}{4X_\nu}
    {+}\frac{\tilde \Psi_\nu}{X_\nu}\iota_{\lst{\nu}}
    {+}\frac{\eps}{2x_\nu}\Bigr)\sigma_{\lst{\nu}} \\
&\mspace{100mu}- \eps\, \frac{i\sqrt{-c}}{2x_\nu^2} -\chf{\nu}\Biggr]\,\bigotimes_\kappa \chi_\kappa=0\,.
\end{split}\raisetag{5ex}
\end{equation}
We finally note that the operators act only on the $\chi_\nu$ spinor, leaving invariant all the other spinors in the tensor product. So we are left with the following ordinary differential equation for each spinor $\chi_\nu$:
\begin{equation}\label{chieq}
\begin{split}
&\Biggl[\Bigl(
    \frac{d}{dx_\nu}+\frac{X_\nu'}{4X_\nu}
    +\frac{\tilde \Psi_\nu}{X_\nu}\iota_{\lst{\nu}}
    +\frac{\eps}{2x_\nu}\Bigr)\,\sigma_{\lst{\nu}} \\
&\mspace{70mu}- \,\frac{\bigl(- \iota_{\lst{\nu}} \bigr)^{\!N\!-\!\nu}}{\sqrt{|X_\nu|}}
   \Bigl(\eps \frac{i\sqrt{-c}}{2x_\nu^2} +\chf{\nu}\Bigr)\Biggr]\,\chi_\nu=0\,.
\end{split}
\end{equation}
This final equation is of the same form as for the uncharged case \cite{CarigliaEtal:2011b}; the only difference is the appereance of an extra term proportional to the coupling constant $e$ in
\eqref{psfdef} and \eqref{chiN}.

As discussed in \cite{CarigliaEtal:2011b} one can, by using the unitary transformation which makes the $\gamma$ matrices coordinate dependent
\be
\tilde \gamma^a=R^{-1}\gamma^a R\,,
\ee
achieve the standard separability (without prefactor $R$) in the form
\begin{equation}\label{tens_sep}
    \psi = \exp\bigl({\textstyle i\sum_k\psc{k}\psi_{k}}\bigr)\,
           \bigotimes_\nu \chi_\nu\;,
\end{equation}
where $\chi_\nu$ satisfy the equation \eqref{chieq}.
The separated solution is an eigenfuction of operators $\tilde M_j\equiv R^{-1}M_j R$ and $K_k$, i.e.,
a solution of
\be\label{eigenvaluetilde}
K_k \psi = i\,\psc{k}\psi\;,\quad \tilde M_j \psi = \chc{j}\psi\;.
\ee

To conclude, we have demonstrated complete separation of variables for a charged Dirac equation in a weakly charged black hole spacetimes in all dimensions. Separated
equations \eqref{chieq} can be in principle decoupled, to yield a second order ordinary differential equation for each component of a 2-dimensional spinor $\chi_\nu$.
It is easy to see that if instead of weakly charged black holes we considered weakly magnetized black holes the demonstrated procedure leading to the separated equation \eqref{chieq} would break down at several places and hence the separability of the charged Dirac equation in this case is not possible, at least in the `canonical setup'.
We expect this to be true in any reference frame.

\section{Summary}\label{sec7}
Kerr-NUT-(A)dS rotating black holes in arbitrary dimension are distinguished by the remarkable fact that several equations of physics in such a background can be solved by a separation of variables. This includes, among others, the Hamilton--Jacobi, Klein--Gordon and Dirac equations. Ultimately, the reason for this is the presence of a PKY tensor. It is possible to consider perturbations of these metrics by turning on an electromagnetic gauge potential along the directions of the Killing vectors. As it was shown in \cite{FrolovKrtous:2011}, among these only the perturbation that corresponds to a weakly charged black hole is such that it is compatible with separability of the Hamilton--Jacobi and Klein--Gordon equations.

In this work we have studied the perturbed metrics from the point of view of the Dirac equation, analysing the presence of symmetry operators of the Dirac equation and their relation to separability. A discussion of first order symmetry operators for the Dirac equation in the presence of an arbitrary flux can be found in \cite{KubiznakEtal:2010}. In particular, it is shown there how in order to build symmetry operators for the Dirac equation in the presence of $U(1)$ flux it is necessary to both i) have Killing--Yano tensors and ii) satisfy additional conditions of absence of anomalies. For all the perturbed
Kerr-NUT metrics there is a candidate complete set of first order symmetry operators. However, we have proved that only for the weakly charged solution there are no anomalies. For this solution we have shown that all such operators mutually commute. We have also explicitly expressed the symmetry operators in a chosen frame, adapted to the hidden symmetry present in the spacetime, and demonstrated that in such a frame the Dirac equation admits a solution by separating variables. These results generalise those found in \cite{FrolovKrtous:2011} and \cite{OotaYasui:2008, CarigliaEtal:2011a, CarigliaEtal:2011b}.


\section*{Acknowledgments}
V.F. thanks the Natural Sciences and Engineering Research Council of Canada and the Killam Trust for the financial support.
P.K. was supported by Grants GA\v{C}R~202/09/0772 and GA\v{C}R~P203/12/0118 and
acknowledges hospitality at the University of Alberta where this work was partially done.
M.C. was supported by Fapemig under the project CEX APQ 2324-11.

\appendix

\section{Important identity}\label{appA}
In this Appendix we prove the identity \eqref{Faeheh+=0}. We start with some auxiliary definitions. Let ${\hdotp{k}}$ be the ${k}$-th ``matrix'' power of the PKY tensor ${h}$,
\begin{equation}\label{hdotpow}
    \hdotp{k}{}_{ab} = h{}_{aa_2}h^{a_2}{}_{a_3}\dots h^{a_k}{}_b
\end{equation}
For odd ${k}$, it is antisymmetric 2-form. (Note, ${\hdotp{k}}$ is different from ${\CCKY{k}}$!) Since it is both Lie and parallel conserved along ${\PKV}$, we have
\begin{equation}\label{dxihdotp=0}
    d\PKV^\flat\cwedge{1}\hdotp{2j+1} = 0\;.
\end{equation}
We also introduce 1-form with one vector index
\begin{equation}\label{omdef}
    \omh{j}^a=X^a\hook\hdotp{j}\;,
\end{equation}
which satisfies
\begin{equation}\label{omhrel}
    \omh{j}^a{}^{\!\sharp}\hook\CCKY{k} = \omh{j+1}^a\wedge\CCKY{k-1}\,.
\end{equation}

In the following we will need also two standard identities for the contracted wedge product involving 1-form~${\omega}$, ${p}$-form ${\alpha}$ and ${q}$-form ${\beta}$:
\begin{equation}\label{cwedgid}
\begin{gathered}
  (\omega\wedge\alpha)\cwedge{m}\beta = (-1)^m \omega\wedge(\alpha\cwedge{m}\beta)
    + m\,\alpha\cwedge{m{-}1}(\omega^\sharp\hook\beta)\;,\\
  \alpha\cwedge{m}(\omega\wedge\beta) = (-1)^p \omega\wedge(\alpha\cwedge{m}\beta)
    + m\,(\omega^\sharp\hook\alpha)\cwedge{m{-}1}\beta\;.
\end{gathered}
\end{equation}
The identity \eqref{hh=0} is based on the relation
\begin{equation}\label{hcwoh=0}
  \CCKY{k}\cwedge{2m+1}\CCKY{l} = 0\;,
\end{equation}
see the proof in \cite{CarigliaEtal:2011a}.

Now we can proceed in our proof of \eqref{Faeheh+=0}. First, we show the identity
\begin{equation}\label{Fomomhh=0}
  (\nabla_{\![a}\PKV_{b]})\; \omh{2j}^a\wedge
    \Bigl(\bigr(\omh{2j}^b{}^{\!\sharp}\hook\CCKY{k}\bigr)\cwedge{2m}\CCKY{l}\Bigr)=0\;.
\end{equation}
Expanding the mixed product with the help of \eqref{cwedgid}, using \eqref{omhrel} and the identity \eqref{hcwoh=0}, we obtain
\begin{equation}\label{derl1}
\begin{split}
  &\bigr(\omh{2j}^b{}^{\!\sharp}\hook\CCKY{k}\bigr)\cwedge{2m}\CCKY{l}
    = \omh{2j+1}^b\wedge\bigl(\CCKY{k-1}{\cwedge{2m}}\CCKY{l-1}\bigr)\\
  &\quad    + (2m)(2m{-}1) \bigl(\omh{2j+2}^b{}^{\!\sharp}\hook\CCKY{k-1}\bigr)\cwedge{2m-2}\CCKY{l-1}\;.
\end{split}
\end{equation}
Iterating this relation, we find
\begin{equation}\label{derl2}
\begin{split}
  &\bigr(\omh{2j}^b{}^{\!\sharp}\hook\CCKY{k}\bigr)\cwedge{2m}\CCKY{l}\\
  &\quad  = \sum_{i=0}^m \frac{(2m)!}{(2m{-}2i)!}\,
    \omh{2j+2i+1}^b\wedge\bigl(\CCKY{k-i}\cwedge{2m-2i}\CCKY{l-i}\bigr)\;.
\end{split}\raisetag{9ex}
\end{equation}
Substituting back to \eqref{Fomomhh=0} we obtain a sum of terms, each of which contains the expression\footnote{%
Here ``${\cdot}$'' indicates the contraction, i.e.,
${(\hdotp{2j}\cdot(\dots)\cdot\hdotp{2j})_{ab}}={\hdotp{2j}{}_a{}^c\,(\dots)_{cd}\,\hdotp{2j}{}^d{}_b}$.}
\begin{equation}\label{Foo}
    \bigl(\nabla_{\![a}\PKV_{b]}\bigr)\,\omh{2j}^a\wedge\omh{2j+2i+1}^b
      = \hdotp{2j}\cdot\bigl(d\PKV^\flat\cwedge{1}\hdotp{2i+1}\bigr)\cdot\hdotp{2j}\,,
\end{equation}
which vanishes thanks to relation \eqref{dxihdotp=0}. We thus proved the relation \eqref{Fomomhh=0}.

Next, we prove that
\begin{equation}\label{Fehcwoeh=0}
  \bigl(\nabla_{\![a}\PKV_{b]}\bigr)\;
   \bigl(e^a\wedge\CCKY{k}\bigr)\cwedge{2m+1}\bigl(e^b\wedge\CCKY{l}\bigr)=0\;.
\end{equation}
Employing again the relations \eqref{cwedgid} for mixed products, the identity \eqref{hcwoh=0}, the just proven relations \eqref{Fomomhh=0}, the antisymmetry of ${\nabla_{\![a}\PKV_{b]}}$, and relation \eqref{omhrel}, we find
\begin{equation}\label{derL2}
\begin{split}
  &\bigl(\nabla_{\![a}\PKV_{b]}\bigr)\;
  \bigl(e^a\wedge\CCKY{k}\bigr)\cwedge{2m+1}\bigl(e^b\wedge\CCKY{l}\bigr)\\
  &\;= (2m{+}1)(2m) \bigl(\nabla_{\![a}\PKV_{b]}\bigr)\;
  \bigl(\omh{2}^a\wedge\CCKY{k}\bigr)\cwedge{2m-1}\bigl(\omh{2}^b\wedge\CCKY{l}\bigr)\,.
\end{split}\raisetag{8ex}
\end{equation}
Defining ${e^a=\omh{0}^a}$, this expression suggests the recurrence relation between ${\omh{2j}}$ and ${\omh{2j{+}2}}$, which can be proved by the same procedure. Iterating it ${(m+1)}$-times, in the last step the expression annihilates completely.

Finally, the identity \eqref{Fehcwoeh=0} implies the desired relation \eqref{Faeheh+=0}. Indeed, using the standard relations for Clifford anticommutator (see, e.g., (A19) of \cite{CarigliaEtal:2011a}), we get
\begin{equation}\label{proof}
\begin{split}
  &(\nabla_{\![a}\PKV_{b]})\;\bigl[ e^{[a}\wedge\CCKY{k},\,e^{b]}\wedge \CCKY{l}\bigr]_+\\
  &\;= \sum_m \frac{2(-1)^m}{(2m{+}1)!}  \bigl(\nabla_{\![a}\PKV_{b]}\bigr)
  \bigl(e^{a}\wedge\CCKY{k}\bigr)\cwedge{2m{+}1}\bigl(e^{b}\wedge \CCKY{l}\bigr) = 0\;.
\end{split}\raisetag{9ex}
\end{equation}

\section{Direct proof of mutual commutativity of operators $M_j$} \label{appB}
In this section we prove mutual commutativity of operators $M_j$, using their explicit representation \eqref{Mopexpl}; the presentation closely follows Sec.~VI.A in \cite{CarigliaEtal:2011b}.
Let us start from the expression for $M_j$ \eqref{Mopexpl}  and apply the identity  \eqref{Sid}, together with
\ba
 \sum_\mu \frac{\Bo{j}_\mu}{{\iota_{\lst{\mu}}x_\mu} \Vo_{\mu}}&=&\frac{{\Bo{j}}}{\Bo{N}}\,,\label{eq:Bj_over_Bn}\\
\Bo{j-1}=\Bo{j}\sum_\mu \frac{1}{\iota_{\lst{\mu}}x_\mu}&-&\Bo{N}\sum_\mu\frac{\Bo{j}_\mu}{\Vo_{\mu}x_\mu^2}\,,\label{id15}
\ea
to obtain
\begin{equation}\label{Mopexpl2}
\begin{split}
M_j
&= i^j \!\sum_\mu\! \sqrt{Q_\mu} \Bo{j}_\mu\Biggl(
    \cd{x_\mu}+\frac{X_\mu'}{4X_\mu} +  \frac12 \sum_{\substack{\nu\\\nu\neq\mu}} \frac1{x_\mu\!-\!\iota_{\lst{\mu\nu}}x_\nu}  \\
-& \frac{i\iota_{\lst{\mu}}}{X_\mu}\!\sum_k (-x_\mu^2)^{N{-}1{-}k}\cd{\psi_k}
    +\frac{\eps}{2x_\mu} - e\iota_{\lst{\mu}}\Biggr)\gamma^\mu   \\
-& \frac{1}{2}\eps\, i^{j+1} \sqrt{-c} \sum_\mu\frac{\Bo{j}_\mu}{\Vo_\mu}\left[\frac{1}{x_\mu^2}
+\frac{1}{\iota_{\lst{\mu}} x_\mu} \left( \frac{2}{ic} \cd{\psi_N} + 2e \right) \right] \, .
\end{split}\raisetag{13.5ex}
\end{equation}
In order to prove commutativity of these operators we introduce new `auxiliary' operators
\be\label{RMR}
\tilde M_j\equiv R^{-1}M_jR\,,
\ee
with $R$ given by \eqref{Phidef}.
Then, obviously, if
\be\label{com}
[\tilde M_j, \tilde M_k]=R^{-1}[M_j,M_k]R=0\,,
\ee
then the same is true for the operators ${M_j}$.
We calculate
\begin{equation}\label{Mopexpl3}
\begin{split}
&M_jR
= i^j \!\sum_\mu\! \sqrt{Q_\mu} \Bo{j}_\mu R R^{-1}\gamma^\mu R\,\times\\
&\times \Biggl(
    \cd{x_\mu}\!+\!\frac{X_\mu'}{4X_\mu}\!+\!\frac{\eps}{2x_\mu}
    \!+\! \frac{i\iota_{\lst{\mu}}}{X_\mu}\!\sum_k (-x_\mu^2)^{N\!-\!1\!-\!k}\cd{\psi_k}  \!+\! e\iota_{\lst{\mu}}
    \Biggr)   \\
&- \frac{1}{2}\eps\, i^{j+1} \sqrt{\!-c} \sum_\mu\frac{\Bo{j}_\mu}{\Vo_\mu}\!\left[\frac{1}{x_\mu^2}
\!+\!\frac{1}{\iota_{\lst{\mu}} x_\mu} \left( \frac{2}{ic} \cd{\psi_N} \!+\! 2e \right) \right]R \, .
\end{split}\raisetag{21.5ex}
\end{equation}
where  we have used \eqref{derPhi}.
The $R$ factor in the $\varepsilon$ term on the right hand side can be brought to the front whereas for the product $R^{-1}\gamma^\mu R$ we use \eqref{eq:gamma_mu_phi}.
So we get
\be \label{tildeMj}
\tilde M_j =i^j\sum_\mu \frac{\Bo{j}_\mu}{ \Vo_{\mu}}\tilde M_\mu \,,
\ee
where the operators
\ba
\tilde M_\mu&=&\sqrt{|X_\mu|}\Biggl(
    \cd{x_\mu}+\frac14\frac{X_\mu'}{X_\mu} +  \frac{\eps}{2x_\mu}   - e\iota_{\lst{\mu}}\nn\\
      &-&   \frac{i\iota_{\lst{\mu}}}{X_\mu}\!\sum_k (-x_\mu^2)^{N{-}1{-}k}\cd{\psi_k} \Biggr)
\left( - \iota_{\lst{\mu}}\right)^{N-\mu} \sigma_{\lst{\mu}}\,,\nn\\
&-& \frac{i}{2}\eps\,\sqrt{-c}\left[\frac{1}{x_\mu^2}
+ \frac{1}{\iota_{\lst{\mu}} x_\mu} \left( \frac{2}{ic}\cd{\psi_N} + 2e \right) \right] \,
\ea
act only on spinor $\chi_\mu$ and hence $[\tilde M_\mu,\tilde M_\nu]=0$.
Using \eqref{BVrel} we can invert the relation \eqref{tildeMj},
\be\label{tildeMmu}
\tilde M_\mu=\sum_{j=0}^{N-1}(-i)^j(-\iota_{\lst{\mu}}x_\mu)^{N-1-j}\tilde M_j\,.
\ee
Following now the procedure adopted in \cite{SergyeyevKrtous:2008}, and using that
$[\tilde M_\mu, (-\iota_{\lst{\nu}}x_\nu)^{N-1-j}]=0$, we establish that
\ba
&&\hspace{-0.5cm}\sum_{j,k=0}^{N-1}(-i)^{j+k}(-\iota_{\lst{\mu}}x_\mu)^{N-1-j}(-\iota_{\lst{\nu}}x_\nu)^{N-1-k}\times\nonumber\\
&&\hspace{4cm}\times[\tilde M_j,\tilde M_k]=0\,,
\ea
from which Eq.~\eqref{com} follows.





\providecommand{\href}[2]{#2}\begingroup\raggedright\endgroup

\end{document}